\documentclass[pre,preprint]{revtex4}
\def\po{pseudotrajectory }
\def\pos{pseudotrajectories }
\def\poe{pseudotrajectory}
\def\pose{pseudotrajectories}
\def\ka{\tau}
\def\ep{\tilde{\epsilon}}
\def\err{\epsilon}
\usepackage{graphics}
\begin{document}

\title{Sampling Chaotic Trajectories Quickly in Parallel}
\author{J. Machta}
\email[]{machta@physics.umass.edu}
\address{Department of Physics,
University of Massachusetts,
Amherst, MA 01003-3720}

\date{\today}

\begin{abstract}
The parallel computational complexity of the quadratic map is studied.  A
parallel algorithm is described that generates typical
pseudotrajectories of length
$t$ in a time that scales as $\log t$ and increases slowly in the
accuracy demanded of the pseudotrajectory.  Long
pseudotrajectories are created in parallel by putting together many short
pseudotrajectories; Monte Carlo procedures are used to eliminate the
discontinuities between these short pseudotrajectories and then suitably
randomize the resulting long pseudotrajectory.  Numerical simulations are
presented that show the scaling properties of the parallel algorithm.  The
existence of the fast parallel algorithm provides a way to formalize the
intuitive notion that chaotic systems do not generate complex histories.
\end{abstract}
\maketitle

\section{Introduction}
\label{sec:intro}
Chaotic systems cannot be predicted for very long times because of the
exponential divergence of nearby trajectories.  Associated with the
divergence of trajectories is a lack of history dependence;
the current behavior of the system is not dependent on the past behavior. 
The absence of history dependence can be understood in various ways.  Here
I take a computational perspective on chaotic systems and analyze
trajectories in terms of computational complexity.  If, using a massively
parallel computer, a typical long trajectory can be manufactured in far
fewer parallel steps than the actual length of the trajectory then the
trajectory lacks history dependence or historical complexity. 
Alternatively, if parallelism does not allow one to generate a typical
trajectory much more quickly than its actual length then the trajectory 
displays a complex history dependence.

These considerations are illustrated using the one-dimensional quadratic
map,
\begin{equation}
x_{n+1} = r x_n(1-x_n) \equiv f(x_n)
\end{equation} 
with $x_n \in [0,1]$ and $0<r<4$.  The objective is to produce typical
trajectories for the map but since computing devices are necessarily
restricted to finite precision we are really interested in generating
typical {\em \pose}~\cite{NuYo}.  A 
\po of accuracy $\delta$ is a
sequence $\{y_n| n=0,\cdots,t\}$ such that
for all $n$ ($0\leq n<t$),
\begin{equation}
\label{eq:po}
|y_{n+1}-f(y_n)|< \delta .
\end{equation}
We suppose that the parallel computer works to a precision
substantially better than $\delta$ so that bounds such as given in Eq.\
\ref{eq:po} can be checked with reasonable certainty. 

For chaotic dynamics
with a positive Lyapunov exponent
$\lambda$, a
\po and an exact trajectory that are initially equal remain
close only for a time that is roughly given by
$(\log\delta)/\lambda$. Nonetheless, for small $\delta$ a \po will have
nearly the same statistical properties as a real trajectory and, in any
case,  numerical results about chaotic systems are learned from \pos not
real trajectories. 

The goal then is to produce a \po chosen from
the uniform distribution over \pose.  A typical \po of length $t$ can be
generated using one processor in time linear in $t$ by iterating the
map using arithmetic of precision much better than
$\delta$ and then adding a noise term on each step chosen from the
uniform distribution on $[-\delta,\delta]$.  Could a typical trajectory be
produced in far fewer than
$t$ parallel steps  with the aid of many processor?  

The model of parallel computation implicit in this
question is the PRAM (parallel random access machine), the standard
model in the theory of parallel computational complexity~\cite{GrHoRu}.  A
PRAM is an idealized, fully scalable device with many identical (except for
distinct integer labels) processors.  Processor all run the same program
and all communicate with a global memory in unit time.  Massive
parallelism is envisioned here however the number of processors
is required to be polynomially bounded in 
$t \log (1/\delta)$, the effective number of (binary) degrees of freedom of
a \po of length $t$ and accuracy $\delta$.

 In the next section we
show how to produce a typical trajectory in parallel using Monte Carlo path sampling.  The procedure is
correct but inefficient.  We then describe how simulated annealing
together with path sampling can produce a typical \po in
parallel time that scales linearly in $\log t$
and polynomially in $\log (1/\delta)$ using a number of
processors that is polynomial in
$t \log (1/\delta)$.

\section{Path sampling of pseudotrajectories}
\label{sec:ps}

The method described in this section is based on path sampling ideas put
forward by Chandler and collaborators~\cite{DeBoCsCh}.  The uniform
probability density for \pos, ${\cal P}(y_0,y_1, \ldots , y_t)$ is given
by 
\begin{equation}
\label{eq:typ}
{\cal P}(y_0,y_1, \ldots , y_t)= p(y_0) \prod_{n=1}^t P(y_n | y_{n-1})
\end{equation}
where
\begin{equation}
\label{eq:typ1}
P(y^\prime |y)=\left\{ \begin{array}{ll}
		\frac{1}{2\delta} & \mbox{if $|y^\prime-f(y)|< \delta$}\\
		0 & \mbox{otherwise}
\end{array}
\right.
\end{equation} 
and $p(y)$ is the invariant distribution.

A simple Monte Carlo procedure can be used to sample paths from ${\cal
P}$.  Given an existing \poe, a single time $n>0$ is chosen and a 
proposal for a new value for
$y_n$ is obtained according to
\begin{equation}
\label{eq:prop}
y_n^\prime=f(y_{n-1})+\epsilon
\end{equation}
where $\epsilon$ is chosen from the uniform distribution on
$[-\delta,\delta]$. The proposed value is accepted if it is also the
case that $|y_{n+1}-f(y_n^\prime)|< \delta$.  It is straightfoward to
verify that this Monte Carlo procedure satisfies detailed balance with
respect to ${\cal P}$.  The question of ergodicity of the Markov chain in
the space \pos is less clear.  However, even if ergodicity holds, the
actual mixing time for the Monte Carlo procedure would be long when
$\delta$ is small since the time to obtain an independent \po is at
least as great as $1/\delta^2$, the time to diffuse a distance order one
given a step size of $\delta$.  To be considered an efficient process for
generating \pos, the parallel time should increase no more rapidly than some
power of the {\em logarithm} of $1/\delta$.  This goal can be achieved by
first using a simulated annealing procedure to produce a \po that is random
on long time scales then to use the above Monte Carlo path sampling to 
randomize the small scales.

\section{Simulated annealing for pseudotrajectories}
\label{sec:an}
Long \pos are constructed by independently generating many {\em
segments} or short trajectories and then ``welding'' the segments together. 
In the welding step the discontinuity between successive segments is
eliminated by simulated annealing. In
addition to simulated annealing, it is sometimes necessary to extend 
segments to obtain a weld to the next segment. 

The fundamental time scale in the system is
\begin{equation}
\ka=-(\log\delta)/|\lambda| ,
\end{equation}
the time required for typical errors to grow to be order one.
The length of the segments, $K$ used in the construction should be longer
than the fundamental  time $\tau$ so that the beginning and end of each
segment is uncorrelated.   We also need to allow for extensions at either
end of the segment and for a ``warm-up'' so that the initial point of the
segment is chosen from invariant distribution.  Thus, in practice, for
each segment we start with a random number and iterate the map $L>K$ times,
choosing the segment of length $K$ from a predetermined part of the longer
sequence of length $L$. 
 
Having made a collection of segments, we now attempt to
weld them together into a long \po.  This is done in such a way that the
initial value or {\em anchor point} of each segment is held fixed.  The
discontinuity between successive segments is
annealed until all  errors are less than
$\delta$.  The Monte Carlo annealing procedure is designed to lower the
error
$e(y_{n-1},y_n,y_{n+1})$ associated with three successive
elements, $y_{n-1},y_n$ and $y_{n+1}$,
\begin{equation}
\label{eq:defe}
e(y_{n-1},y_n,y_{n+1})= |y_n-f(y_{n-1})|+|y_{n+1}-f(y_n)|   .
\end{equation}
If $e(y_{n-1},y_n,y_{n+1}) < \delta$ for every $n$ then we have a \poe. 
For each time $n$, with the exception of the anchor points, the Monte Carlo
annealing procedure begins with a measurement of
$e=e(y_{n-1},y_n,y_{n+1})$.  If $e<\delta$ nothing is done.  Otherwise
a new value
$y_n^\prime$ is proposed,
\begin{equation}
y_n^\prime= y_n + \ep
\end{equation}
where $\ep$ is chosen as a Gaussian random variable with mean zero
and standard deviation $e/2$.  If $e^\prime=e(y_{n-1},y_n^\prime,y_{n+1})$,
is less than $e$, the proposal is accepted as the new value for $y_n$. 
If the error increases, the proposal is accepted with probability
$e^{-\beta(e^\prime-e)}$.  The value of the inverse temperature for each
Monte Carlo step and is taken to be
$\beta=e/2$ so that the acceptance ratio is independent of the size
of the error.  In a single Monte Carlo sweep, all except the initial and
final values of each segment are processed using the above procedure.  Given
$t/2$ processors this can be done in constant parallel time by first
processing the even and then the odd values of the time $n$.

In the chaotic regime ($\lambda>0$) the annealing
procedure should yield a valid \po for large enough $K$ and sufficiently
many Monte Carlo sweeps.  In practice, however, some welds require
a very large number of sweeps.  Specifically, the probability distribution
for the number of sweeps needed to achieve a weld has a long tail leading
to parallel running times for creating \pos that are dominated by the few
most difficult welds.  Two additional kinds of steps, called  {\em forward
shifts} and {\em backward shifts} cure this difficulty.  Suppose
segment
$s$ together with its final condition, the first element of segment $s+1$,
is not fully annealed after a predetermined number of annealing sweeps. Then
segment $s$ is restored to its original state and either it is
extended forward one step or segment
$s+1$ is extended backwards one step.  In the case of a backwards
shift, the element prepended to segment $s+1$ is considered a new anchor
points and serves as the new final condition for segment $s$.  The net
effect of either a forward or backward shift is that the discontinuity
between segments
$s$ and $s+1$ occurs with a different pair of numbers. Annealing sweeps and
shifts are interleaved, a fixed number of Monte Carlo annealing sweeps
are attempted and if all errors are not less than $\delta$, a shift is
done.  The process is repeated until a 
satisfactory weld is achieved.  Successive shifts are alternately of the
forward and backward type. In Sec.\
\ref{sec:vs}, I show that the combination of Monte Carlo annealing and
shifts produces a \po in $O(\log t)$ parallel steps.  

Shifts serve several purpose.  First, they simply provide for the
possibility of more Monte Carlo sweeps, though if this were their only
function it could be accomplished by directly increasing the number of
sweeps.  Second, shifts permit the algorithm to perform properly for
periodic orbits or nearly periodic stretches of aperiodic orbits. The
annealing procedure by itself cannot generate long 
\pos for  periodic orbits since welds are often 
attempted between segments that are out of phase with one another. Adding
shifts to the annealing procedure insures that periodic \pos will be
correctly generated.  For example, consider the case of a period two
orbit, a single forward or backward shift of some segments will
insure that all welds are satisfactory.  For period
$d$ orbits, as many as $d-1$ shifts are necessary to insure that all
welds are satisfactory. 

Shifts may also provide padding around hard to weld regions of a trajectory.
During a shift, new points are added to one end of a segment
but no points are removed.  Thus shifts do not bias the \po against
difficult to weld regions in the invariant measure.  For example, it is
observed that if the final condition for a segment is very near the maximum
of the support in the invariant measure at
$r/4$ then one or more backward shifts are usually necessary so that the
point near $r/4$ is surrounded by a region of small errors and is not
involved in the annealing process.

\section{Full parallel algorithm for pseudotrajectories}
This section provides the details of the parallel algorithm for producing
typical \pos that combines the path sampling method of Sec.\
\ref{sec:ps} and the annealing procedure of Sec.\ \ref{sec:an}. First the
annealing procedure is used to generate a \po and then path sampling is
used to further randomize it. The algorithm is
controlled by several parameters:
$t$ is the total length of the desired  \po, $\delta$ is the desired
accuracy,
$K$ is the length of each segment, $K^\prime$ is the warm-up length,
$E$, an even number, is the maximum number of shifts that are attempted,
$M_1$ is the number of Monte Carlo annealing sweeps carried out between
shifts and $M_2$ is the number of Monte Carlo path sampling sweeps. 
The algorithm is described below:
\begin{enumerate}

\item
In parallel, generate $S=\lceil t/K \rceil$ sequences $\{x_m^{(s)}\}$
each of length
$L=K^\prime+K+E$,
 with $s=0, \cdots, S-1$ and $m=0, \cdots, L-1$.  The
initial value of each sequence is a uniform random number on $(0,1)$ and
subsequent values are obtained by iterating the map
$L-1$ times to precision much greater than $\delta$.  This step requires
$O(L)$ parallel time.
\item
These $S$ sequences are used to define $ES$ segments $\{y^{(s,q)}_n\}$
each of length $K$,  where the index $q$,
$0 \leq q < E$ gives the number of shifts. $y^{(s,q)}_n =
x^s_{n+E/2+K^\prime+\lfloor q/2 \rfloor}$ for $0 \leq n<K-1$
while $y^{(s,q)}_{K-1} = x^{s+1}_{E/2+K^\prime-\lceil q/2
\rceil+K^\prime}$.  Note that the final point in segment $y^{(s,q)}$ is a
taken from sequence $s+1$.
\item
In parallel, for each $s<S$, and each $q<E$,  anneal segment
$y^{(s,q)}$.  The annealing procedure consists of
$M_1$ Monte Carlo sweeps.  During a single annealing sweep first the even
 and then the odd elements of
the segment are updated in parallel.  The initial and final points
$y_0^{(s,q)}$ and $y_{K-1}^{(s,q)}$ are held fixed during the annealing
procedure.  A single Monte Carlo update of the point
$y^{(s,q)}_m$ consists of the following procedure:
\begin{itemize}
\item
Compute $e=e(y^{(s,q)}_{m-1},y^{(s,q)}_m,y^{(s,q)}_{m+1})$ from Eq.\
\ref{eq:defe}.
\item
If $e<\delta$, do nothing. If $e > \delta$ propose a new value $y^\prime
=y^{(s,q)}_m + \ep$ where $\ep$ is a Gaussian random variable
with mean zero and standard deviation $e/2$.
\item
Compute $e^\prime= e(y^{(s,q)}_{m-1},y^\prime,y^{(s,q)}_{m+1})$.  If
$e^\prime \leq e$ accept the proposed change, $y^{(s,q)}_m \leftarrow
y^\prime$.  If $e^\prime > e$ accept the proposed move with probability
$\exp[-\beta(e^\prime-e)]$ where $\beta=e/2$. 
\end{itemize} 
The parallel time required for this step is $O(M_1)$.
\item
In parallel, for each $s<S$ find $Q(s)$, that smallest value of $q$
such that the segment is successfully annealed.  The annealing is successful
for this segment if, for all
$0<m<K-1$, the errors are sufficiently small,
$e(y^{(s,q)}_{m-1},y^{(s,q)}_m,y^{(s,q)}_{m+1}) < \delta$.  If for
any $s$, annealing is unsuccessful for all $q \leq E$, the
algorithm fails. This step can be carried out in constant parallel time.
\item
The full \po $y^*$ is a concatenation of sequences obtained from each
original sequence
$x^{(s)}$.  The contribution to the \po from $x^{(s)}$ is the
concatenation of $\{x_m^{(s)} | E/2+K^\prime-\lceil Q(s-1)/2 \rceil \leq m
\leq E/2+K^\prime+\lfloor Q(s)/2 \rfloor\}$ and
$\{y_m^{(s,Q(s))}|0 < m< K-1 \}$.  The first of these sequences is
composed of the anchor points and the second sequence is the annealed
segment.  To obtain a 
\po of length exactly
$t$, the
\po obtained above is simply truncated after $t$ steps.
\item
The path sampling Monte Carlo procedure described in Sec.\ \ref{sec:ps}
further randomizes 
${y^*_m}$. During a single sweep, first all the even  and then all the odd
elements of the \po are updated in parallel.  The number of sweeps is
$M_2$.  On each Monte Carlo step  a  new value for $y^*_m$ is proposed
according to Eq.\ \ref{eq:prop} and accepted only if the trajectory is
still a
\po within error
$\delta$.  The randomization step requires parallel time $O(M_2)$.

\end{enumerate}

\section{Validity and complexity of the parallel algorithm}
\label{sec:vs}

The central questions addressed in
this section are (1) whether the algorithm succeeds in creating a
\poe, (2) how the scaling of the number of
parallel steps depends on the length and accuracy of the \po and the
parameter $r$ of the map and (3) whether the algorithm samples the uniform
distribution on \pose.  A sequential algorithm that carries out the
annealing and path sampling routines one segment at a time was used to
study these questions. In the simulations reported below, the parameters
are chosen to be
$M_1=M_2= 5
\tau^2$, $K= 5
\tau$ and
$K^\prime =1000$.  The assumption behind these choices is that memory is
lost on a time scale $\tau$ so that placing independently chosen anchor
points separated by
$K=5 \tau$ is satisfactory.  The annealing process that welds
successive segments is expected to influence a region whose length
is order $\tau$. Since information is transmitted diffusively by
local Monte Carlo moves, having the number of Monte Carlo sweeps
scale as $\tau^2$ should suffice.  

\begin{figure}
\includegraphics{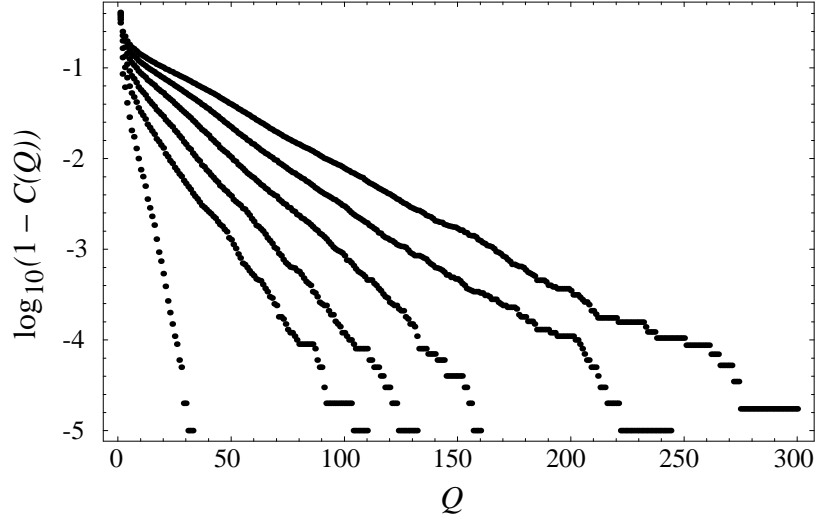}
\caption{Logarithm of the complement of the cumulative distribution for the
number of shifts, $\log_{10}(1-C(Q))$ vs.\ number of shifts, $Q$.  From
left to right,
$\delta=10^{-5}$ to $10^{-10}$.}
\label{fig:cvs}
\end{figure}

The annealing
stage of the parallel algorithm can be studied one segment at a time since
each segment is independently annealed.  First, I observed that, given
enough shifts, the annealing step always produced a successful weld. 
The choice of the maximum number of shifts $E$ for the annealing stage must
be large enough to make the failure probability for the whole algorithm
small. For long trajectories,
$t \gg \tau$, the choice of
$E$ is determined by the tail of the distribution of
the number of shifts, $Q$ required to obtain a weld since the whole
procedure fails if even one segment is not successfully annealed.  Suppose
 $C(\cdot)$ is
the  cumulative probability distribution for $Q$.  An estimate of the
maximum number of shifts,
$E$  needed to insure all segments are annealed is given
by the relation 
\begin{equation}
\label{eq:ests}
(t/K) (1-C(E)) < 1  .
\end{equation}
Figure
\ref{fig:cvs} shows 
$\log_{10}(1-C(Q))$ vs.\ $Q$. These data were collected for the case of
$r=3.7$ (the period doubling transition to chaos occurs at $r=3.5699
\ldots$) and the six curves from left to right are for
$\delta=10^{-5}$ through $10^{-10}$, respectively.  Each curve is obtained
from annealing $10^5$ segments except for the $\delta=10^{-10}$ curve which
is obtained from $6 \times 10^4$ segments.  For $r=3.7$ the Lyapunov
exponent is $\lambda=0.354$ and so, for example, with
$\delta=10^{-7}$, $\tau=45.5$,
$K=228$ and
$M_1=10366$.  Over a reasonable range following an initial transient and
before the noise becomes large, the data falls on straight lines suggesting
that the distribution is asymptotically exponential,
$C(Q)
\sim 1-\exp(-Q/\sigma)$.  Equation \ref{eq:ests} then implies
that $E \sim \sigma \ln t$ and we can conclude that the parallel
running time is $O(\log t)$ since no other contribution to the
running time depends on the overall length of the \poe. 

\begin{figure}
\includegraphics{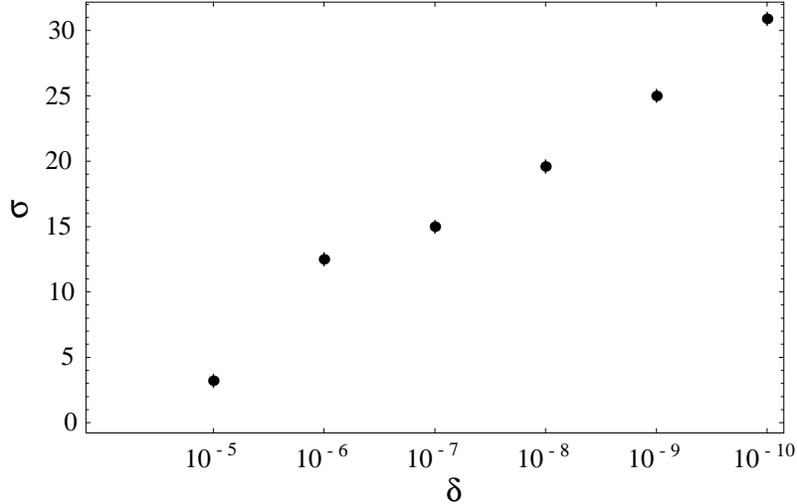}
\caption{Decay constant $\sigma$ for the distribution of shifts vs.\
accuracy $\delta$ on a logarithmic scale.}
\label{fig:svs}
\end{figure}

How does the decay constant $\sigma$ and thus the running time depend on
the choice of the accuracy $\delta$.  Figure \ref{fig:svs} shows $\sigma$
vs.\ $\delta$ on a logarithmic scale for the case
$r=3.7$ and suggests that $\sigma$ is a polynomial
function of
$\log \delta$.  The other simulation parameters, $L$, $M_1$ and $M_2$ are
also polynomial in $\log \delta$ so we conclude that the
full algorithm has a running time that is polynomial in $\log \delta$ and
linear  in $\log t$.  

I also considered two other parameter values for the quadratic map,
$r=3.6$ where $\lambda=0.183$ and $r=3.95$ where $\lambda=0.577$.  In both
cases, the accuracy was set to $\delta=10^{-7}$. The decay of
$(1-C(Q))$ appears to be exponential in both cases but with rather different
values of the decay constant:
$\sigma=450,15$ and $9.5$ for $r=3.6,3.7$ and $3.95$, respectively.  Either
$\sigma$ depends strongly  on $\lambda$ or perhaps there are additional $r$
dependent factors controlling $\sigma$.  For example, for $r=3.7$ and 3.8
the invariant measure has support on a single interval but for $r=3.6$ the
support consists of two intervals.

The annealing stage of the algorithm creates a \po but it is not typical
in the sense of being chosen from the distribution of Eq.\ \ref{eq:typ}. 
On long time scales, the \po is randomized by the random choice
of initial conditions for each sequence. However, the
individual errors, $\err_n=y^*_{n+1}-f(y^*_{n})$ are not
guaranteed to be independent random variables on the interval
$[-\delta,\delta]$.  For example, anchor regions of the \po have errors
much less than $\delta$.  The hypothesis is that
$M_2 = O(
\tau^2)$ path sampling Monte Carlo sweeps are sufficient to randomize the
short time scales and produce a typical
\po from the \po produced by the annealing stage. To check this
hypothesis, I computed mean values and autocorrelation functions for errors
and cross correlations between errors and values of the \po.  The
quantities  $\langle \err_n  \rangle$, ($\langle \err_n^2 \rangle
- \delta^2/3$),
$\langle \err_{n+1} \err_n \rangle$, $\langle \err_{n+1} y^*_n \rangle$
and $(\langle \err_{n+1}^2 \err_n^2 \rangle- \delta^4/9)$ were all found to
be zero within error bars for the case $r=3.7$ and $\delta=10^{-7}$.  Here
the angled brackets indicate an average over segments and over $n$. The
vanishing of these quantities is a necessary but not sufficient condition
that \po is chosen from the uniform distribution described by Eqs.\
\ref{eq:typ} and \ref{eq:typ1}.  More work is needed to
firmly establish that the algorithm with $O(\tau^2)$ path sampling sweeps
samples the uniform distribution to good approximation.  However, even
without the path sampling stage, the \pos produced by the annealing stage
are typical in a different sense. As shown in \cite{NuYo}, any \po {\em
shadows}  an exact trajectory (i.\ e.\ remains close to over its entire
length) though possibly for a larger value of
$r$.
\section{Conclusions}
\label{sec:con}

I have exhibited a parallel algorithm that generates \pos of the quadratic
map. Numerical evidence suggests that the parallel time required to
generate a typical \po increases linearly in $\log t$ and polynomially in
$\log (1/\delta)$ though more work would be required to establish these
scalings with certainty.  Essentially the same parallel algorithm can be
applied to other one-dimensional and higher dimensional maps.  It would be
interesting to explore whether the annealing/shift procedure is sufficient
to efficiently sample \pos for other maps.

Since there is little demand for very long \pos of the quadratic map, the
fast parallel algorithm is probably not of practical value.  The
significance of its existence and complexity is that it
characterizes the history dependence of the map.  The existence of a fast
parallel simulation is a strong statement against history
dependence since it shows that the logical path from independent random
numbers (used to drive the Monte Carlo procedures) to a typical \po
is much shorter than the length of the \poe.  The length of this
logical path is one measure of the potential for generating historical
complexity.  In very few logical steps, very little complexity can be
arise. The idea that complexity tends to emerge at the ``edge of chaos''
\cite{Lang} is born out here since the basic time scale $\tau$ for the
parallel algorithm diverges when the Lyapunov exponent vanishes. An
appealing feature of characterizing systems by computational complexity is
that very different systems in statistical physics systems, for
example diffusion limited aggregation~\cite{MoMaGr97},
sandpiles~\cite{MoNi} or the Bak-Sneppen model~\cite{MaLi01}, can be
compared to one another within the same framework.

\acknowledgements
This work was supported in part by NSF Grant DMR 9978233.

\end{document}